\documentclass[aps,prl,twocolumn,showpacs,superscriptaddress]{revtex4}
\usepackage{amssymb}
\usepackage{amsfonts}
\usepackage{graphicx}
\usepackage{CJK}
\usepackage{indentfirst}
\usepackage{amsmath}
\usepackage{epstopdf}
\begin{document}


\title{Low temperature specific heat of 12442-type KCa$_2$Fe$_4$As$_4$F$_2$ single crystals}

\author{Teng Wang} \affiliation{State Key
Laboratory of Functional Materials for Informatics, Shanghai
Institute of Microsystem and Information Technology, Chinese Academy
of Sciences, Shanghai 200050, China}\affiliation{CAS Center for Excellence in Superconducting
Electronics(CENSE), Shanghai 200050, China}\affiliation{School of Physical Science and Technology, ShanghaiTech University, Shanghai 201210, China}

\author{Jianan Chu}
\affiliation{State Key Laboratory of Functional Materials for
Informatics, Shanghai Institute of Microsystem and Information
Technology, Chinese Academy of Sciences, Shanghai 200050,
China}\affiliation{CAS Center for Excellence in Superconducting
Electronics(CENSE), Shanghai 200050, China}\affiliation{University
of Chinese Academy of Sciences, Beijing 100049, China}

\author{Jiaxin Feng}
\affiliation{State Key Laboratory of Functional Materials for
Informatics, Shanghai Institute of Microsystem and Information
Technology, Chinese Academy of Sciences, Shanghai 200050, China}\affiliation{CAS Center for Excellence in Superconducting
Electronics(CENSE), Shanghai 200050, China}
\affiliation{University of Chinese Academy of Sciences, Beijing 100049, China}

\author{Lingling Wang}
\affiliation{State Key Laboratory of Functional Materials for
Informatics, Shanghai Institute of Microsystem and Information
Technology, Chinese Academy of Sciences, Shanghai 200050,
China}

\author{Xuguang Xu} \affiliation{School of Physical Science and Technology, ShanghaiTech University, Shanghai 201210, China}

\author{Wei Li}\email[]{w_li@fudan.edu.cn}
\affiliation{State Key Laboratory of Surface Physics and Department of Physics, Fudan University, Shanghai 200433, China}
\affiliation{Collaborative Innovation Center of Advanced Microstructures, Nanjing 210093, China}

\author{Hai-Hu Wen}
\affiliation{Center for Superconducting Physics and Materials, National Laboratory of Solid State Microstructures and Department of Physics,
Nanjing University, Nanjing 210093, China}

\author{Xiaosong Liu}
\affiliation{State Key Laboratory of Functional Materials for
Informatics, Shanghai Institute of Microsystem and Information
Technology, Chinese Academy of Sciences, Shanghai 200050, China}\affiliation{CAS Center for Excellence in Superconducting
Electronics(CENSE), Shanghai 200050, China}\affiliation{School of Physical Science and Technology, ShanghaiTech University, Shanghai 201210, China}

\author{Gang Mu}
\email[]{mugang@mail.sim.ac.cn} \affiliation{State Key Laboratory of
Functional Materials for Informatics, Shanghai Institute of
Microsystem and Information Technology, Chinese Academy of Sciences,
Shanghai 200050, China}\affiliation{CAS Center for Excellence in Superconducting
Electronics(CENSE), Shanghai 200050, China}\affiliation{University of Chinese Academy of Sciences, Beijing 100049, China}

\begin{abstract}
Low-temperature specific heat (SH) is measured for the 12442-type
KCa$_2$Fe$_4$As$_4$F$_2$ single crystal under different magnetic
fields. A clear SH jump with the height of $\Delta C/T|_{T_c}$ = 130
mJ/mol K$^2$ is observed at the superconducting transition
temperature $T_c$. It is found that the electronic SH coefficient
$\Delta\gamma (H)$ quickly increases when the field is in the
low-field region below 3 T and then considerably slows down the increase with a further increase in the field, which indicates a rather strong anisotropy
or multi-gap feature with a small minimum in the superconducting
gap(s). The temperature-dependent SH data indicates the presence of the $T^2$ term, which supplies further information and supports the picture with a line-nodal gap structure.
Moreover, the onset point of the SH transition
remains almost unchanged under the field as high as 9 T, which is
similar to that observed in cuprates, and placed this
system in the middle between the BCS limit and the Bose-Einstein
condensation.

\end{abstract}

 \maketitle

\section*{1 Introduction}
From the structural point of view, most Fe-based
superconductors (FeSCs) belong to monolayered, e.g., 1111 and
21311 systems~\cite{LaFeAsO,Zhu2009}, and infinite-layered systems,
e.g., 11 and 122 systems~\cite{FeSe,122}. The only exception is the
12442 system~\cite{12442-1,12442-2,12442-3,12442-4,12442-5}, which
has two FeAs layers between neighboring insulating layers. This
new system has the general chemical formula of AB$_2$Fe$_4$As$_4$C$_2$
(A = K, Rb, and Cs; B = Ca, Nd, Sm, Gd, Tb, Dy, and Ho; C = F and O) and has
attracted considerable research interest in recent
years~\cite{Wang2016,Ishida2017,WangCrystal,WangCrystal-2,WangPauli2020,Kirschner2018,Adroja2018,Smidman2018,Huang2019,Xu2019,Yu2019,WuArpes}.
First principle calculations revealed a rather complicated band
structure with ten Fermi surfaces (FSs)~\cite{Wang2016}. The superconducting (SC) transition
temperature can be tuned by cobalt
substitution~\cite{Ishida2017}. Single crystals with the size of
several millimeters were grown using the self-flux
method~\cite{WangCrystal,WangCrystal-2}. An abnormally high slope of
the upper critical field vs. $T_c$ and a rather strong Pauli
paramagnetic effect have been also
reported~\cite{WangCrystal-2,WangPauli2020}. The SC gap structure
has been investigated by the muon spin relaxation ($\mu$SR), heat
transport, lower critical field, optical spectroscopy, and
angle-resolved photoemission spectroscopy (ARPES) measurements
~\cite{WangCrystal,Kirschner2018,Adroja2018,Smidman2018,Huang2019,Xu2019,
WuArpes}. However, the conclusions are rather controversial. The
lower critical field study revealed a multi-gap feature with a
clear difference in gap sizes~\cite{WangCrystal}. The nodal gap
structure has been indicated by $\mu$SR
measurements~\cite{Kirschner2018,Adroja2018,Smidman2018}. Whereas
the heat transport~\cite{Huang2019}, optical spectroscopy
~\cite{Xu2019}, and recent ARPES
measurements~\cite{WuArpes} have supported a nodeless scenario. Even
within the nodeless scenario, the estimation of the ratio between
large and small gaps ($\Delta_L/\Delta_S$) is also
contradictory. On the basis of heat transport
experiments~\cite{Huang2019}, an estimation $\Delta_L/\Delta_S
\approx 2$ was proposed. The analysis of the lower critical field
data~\cite{WangCrystal} indicates that $\Delta_L/\Delta_S \approx
2.9$. ARPES measurements~\cite{WuArpes} provide an even larger
value of $\Delta_L/\Delta_S \approx 4$. Consequently, more
experiments by other techniques are urgently required to clarify
this argument.

Specific heat (SH) is a bulk tool to detect the quasiparticle density of
states (DOS) at the Fermi level, which can provide information about
the gap structure~\cite{Wen2004,Mu2007,Mu2009,Mu2011}. The variation
in electronic SH in SC states vs. temperature and field is
rather different for different gap
structures~\cite{review1,review2}, which can be used as a reliable
criterion to probe the information of superconducting gap. Although
the SH data have been shown in the previous
studies~\cite{12442-1,Huang2019}, an in-depth investigation is still
lacking. In this study, we present an in-depth low
temperature SH study of the 12442-type KCa$_2$Fe$_4$As$_4$F$_2$
single crystal. A clear SH jump with the height of 130 mJ/mol K$^2$
was observed. A quick increase in the electronic SH coefficient and the presence of the $T^2$ term reveal
a very large anisotropy with line nodes
in the gap structure. Moreover, the feature of the SH anomaly around
$T_c$ under field diverges from the BCS picture and indicates an
evolution tendency toward the Bose-Einstein condensation (BEC).

\section*{2 Materials and methods}
KCa$_2$Fe$_4$As$_4$F$_2$ single crystals were grown by the
self-flux method. The sample for the SH measurement has a mass of
2.3 mg. The detailed growth conditions and sample characterizations
have been reported in our previous work~\cite{WangCrystal-2}. The
magnetic susceptibility measurements were performed with a
superconducting quantum interference device (Quantum Design, MPMS
3). SH was measured on a physical property
measurement system (Quantum Design, PPMS). Similar to that reported in our previous work~\cite{Chu2019}, we employed the thermal
relaxation technique to perform the SH measurements and the
external field was applied along the $c$ axis of the single crystal
during the SH measurements.

The first principles calculations were performed using the
all-electron full potential linear augmented plane wave plus local
orbitals (FP-LAPW+lo) method~\cite{DJSingh}, as implemented in the
WIEN2K code~\cite{PBlaha}. The exchange-correlation potential was
calculated using the generalized gradient approximation as
proposed by Pedrew, Burke, and Ernzerhof~\cite{PBE}. Throughout the
calculations, we fixed the crystal structure to the experimental
values~\cite{12442-1}.

\begin{figure}
\includegraphics[width=8.5cm]{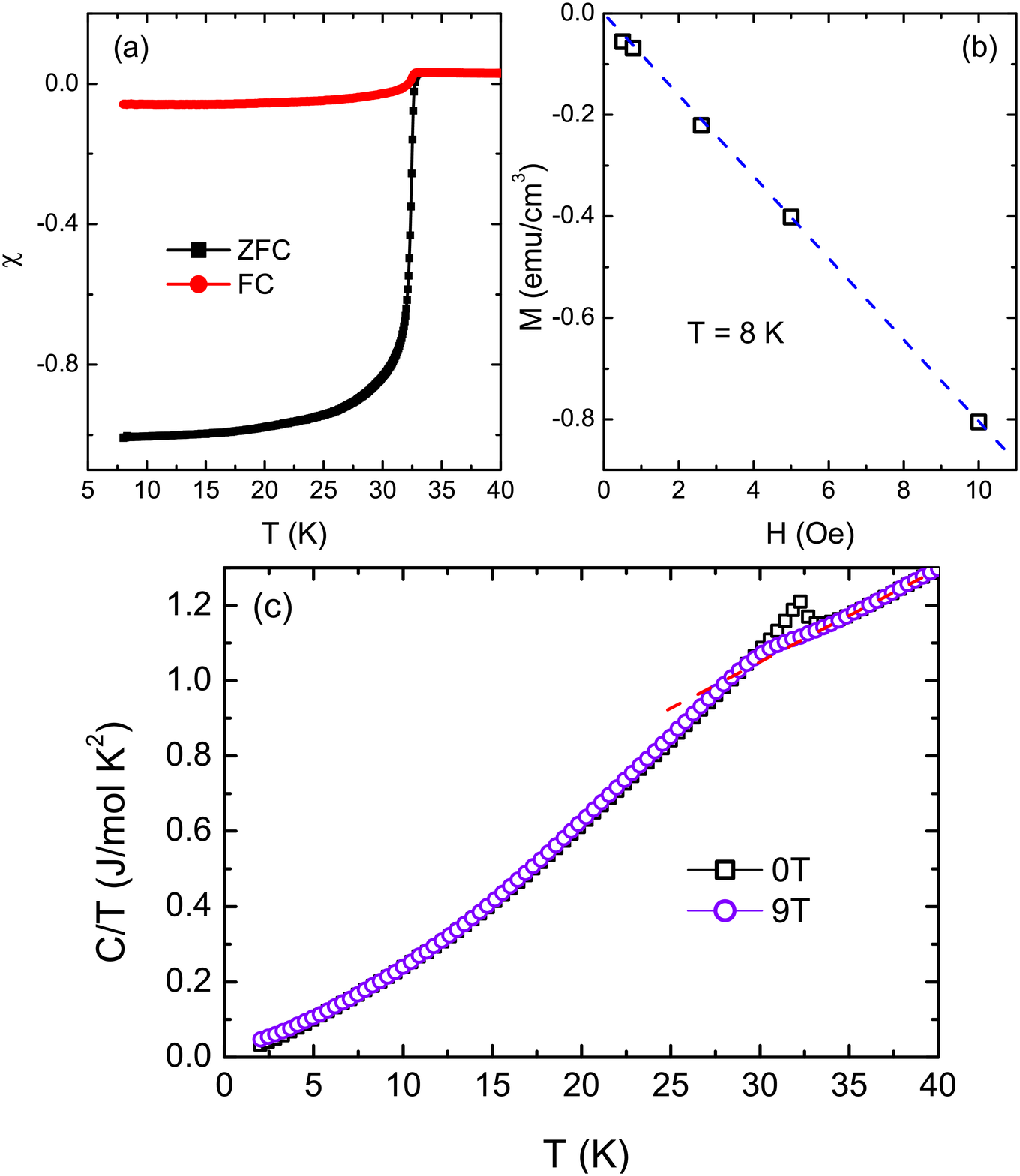}
\caption {(color online) (a) Temperature dependence of magnetic
susceptibility for the KCa$_2$Fe$_4$As$_4$F$_2$ single crystal. The
applied magnetic field is 1 Oe. (b) Field dependence of
magnetization at 8 K. The blue straight dashed line is a guide for
the eye. (c) Temperature dependence of SH plotted as $C/T$
vs. $T$ under the two fields of 0 T and 9 T. } \label{fig1}
\end{figure}

\section*{3 Results and discussion}
The superconducting transition of the single crystal was evaluated by
magnetic susceptibility ($\chi$) measurements. In Fig.
1(a), the $\chi-T$ curve shows a clear and sharp SC transition at approximately 32.5 K, which indicates the high quality of the selected
sample. The magnetic field was applied parallel to the ab-plane of
the crystal to minimize the effect of demagnetization. To ensure the accurate calculation of superconducting volume
fraction, we also measured the field dependence of
magnetization, which is shown in Fig. 2(b). From this
linear in-field behavior, the value of $\chi$ was determined
to be very close to $-1$, which indicated a high superconducting
volume fraction of about 100\% in our sample. Figure 1(c) shows
the raw data for the SH coefficient $\gamma=C/T$ vs. $T$ at 0 T and 9 T.
Here, one mole means the Avogadro number of the formula units (f.u.),
KCa$_2$Fe$_4$As$_4$F$_2$. A clear and sharp SH anomaly owing to the SC
transition can be seen near $T_c$ from the raw data of 0 T. While
for the data under 9 T, this anomaly becomes obscure. Within a
limited temperature range above $T_c$, $C/T$ reveals a linear
temperature dependence, as shown by the red dashed line in Fig. 1(c),
which can be used as an estimation for the SH contribution of the
normal states including the phonon term and the normal electronic
term in this local temperature region in the vicinity of $T_c$.

Thus, we subtracted this linear straight line from the raw data,
and the results are shown in Fig. 2(a). The SH anomaly $\Delta
C/T|_{T_c}$ at zero field was determined to be approximately 130 mJ/mol
K$^2$, as indicated by the blue arrowed line. Of note,
there are 4 Fe atoms in one formula unit of
KCa$_2$Fe$_4$As$_4$F$_2$, which is twice that of 122 system, e.g.
Ba$_{0.6}$K$_{0.4}$Fe$_2$As$_2$. Thus, the SH anomaly has to be
reduced by half when compared with the 122 system, which amounts to
65 mJ/mol K$^2$ for the 2-Fe-atoms case. Assuming a weak-coupling
BCS picture with the ratio $\Delta C/\gamma_n T|_{T_c}$= 1.43, we
can estimate the normal state electronic SH coefficient $\gamma_n
\approx$ 91 mJ/mol K$^2$. For comparison with the
theoretical prediction, we carried out first principles
calculations; the results of electron DOS are shown in Fig. 2(b).
The electronic SH data is related to the value of DOS at the Fermi
level, which is $N(E_F)$ = 4.75 eV$^{-1}$/Fe with both spins
included. This value is quite consistent with the previous report on
this material~\cite{Wang2016} and remarkably larger than that
obtained in 1111 ($\sim$2.62 eV$^{-1}$/Fe) and 122 ($\sim$2.3
eV$^{-1}$/Fe) systems~\cite{Singh2008-1,Singh2008-2}. By disregarding the coupling effects of electrons, the bare electronic SH
coefficient $\gamma_{bare}$ can be calculated using the
formula~\cite{Tari} $\gamma_{bare}=\pi^2k_B^2N(E_F)/3$ = 11.2 mJ/mol
K$^2$ (per mol Fe) = 44.8 mJ/mol K$^2$ (per mol f.u.), where $k_B$
is the Boltzmann constant. This bare value is approximately one half of the
experimental result, which indicates a rather strong coupling of
electrons in the real material.

\begin{figure}
\includegraphics[width=8cm]{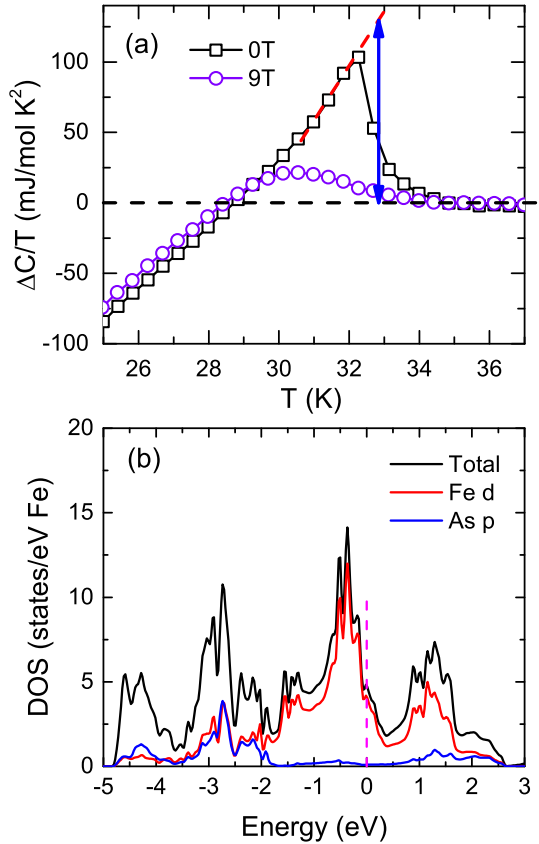}
\caption {(color online) (a) SH data after subtracting the
linear background from the normal states. (b) Calculated electron
DOS of KCa$_2$Fe$_4$As$_4$F$_2$ plotted on a per Fe atom basis with
both spins.} \label{fig2}
\end{figure}

\begin{figure}
\includegraphics[width=8cm]{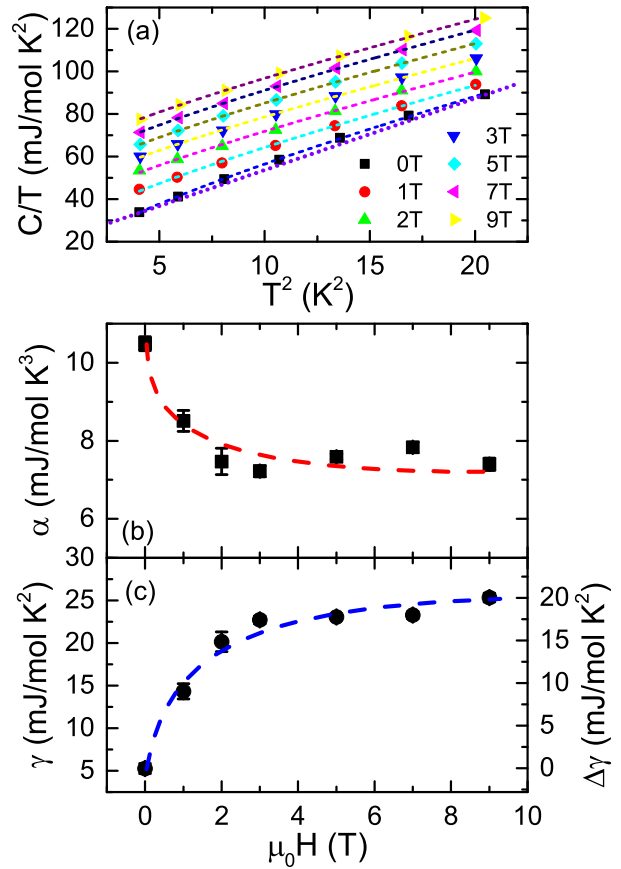}
\caption {(color online) (a) Raw data of SH under different
fields in the low temperature region. The data are shown in the $C/T$ vs.
$T^2$ plot. The data under different fields are shifted upwards by 5 mJ/mol K$^2$ for clarity. The dashed lines are the results of theoretical fitting (see text). The dotted straight line is a guide for the eye.
(b) and (c) Field dependence of the coefficient of the $T^2$ term $\alpha$ and the electronic SH coefficient $\gamma$. The dashed lines are the guide for the eye.}
\label{fig3}
\end{figure}

Next, we focus on the SH data in the low temperature
range below 4.5 K, which is approximately 1/7 of $T_c$, to study low-energy excitations. Here, we plot the raw data
of SH as $C/T$ vs. $T^2$ in Fig. 3(a). The data under different fields are shifted upwards by 5 mJ/mol K$^2$ for clarity. No Schottky anomaly can be
observed, which facilitates the following analysis of our data. As shown by the dotted violet line, all SH data under different
fields reveal a tendency that slightly deviates from linear behavior, which cannot be simply described by
\begin{equation}
C(T,H) = \gamma(H)T+\beta T^3.\label{eq:1}
\end{equation}
Here, the two terms are the normal electronic SH owing to the low-energy excitations by magnetic field and the phonon SH. Similar behaviors were reported in overdoped Ba(Fe$_{1-x}$Co$_{x}$)$_{2}$As$_{2}$ and
Ba(Fe$_{1-x}$Ni$_{x}$)$_{2}$As$_{2}$ systems~\cite{Mu2011,Mu2015}. To effectively simulate the negative curvatures of the experimental data, a $T^2$ term should be considered in the electronic SH:
\begin{equation}
C(T,H) = \gamma(H)T+\alpha(H) T^2 + \beta T^3,\label{eq:2}
\end{equation}
where $\alpha(H)$ is the coefficient of the $T^2$ term under the field
$H$. To obtain reasonable results, during the fitting process, the values of $\beta$ under fields are fixed to that of zero field ($\sim$ 1.8 mJ/mol K$^4$).
As shown by the dashed lines in Fig. 3(a), the fitting result is good. The presence of
the $T^2$ term in electronic SH is the hallmark of the line nodes in the energy
gap(s)~\cite{review1,review2}. The influence of the $T^2$ term on the SH data is more significant at the lower temperatures. Thus, the confirmation by the
lower-temperature experiments is needed in the future. Nevertheless, the obtained result indicates that line nodes are very likely to exist in the gap(s) of the 12442 system,
which is consistent with the $\mu$SR
measurements~\cite{Kirschner2018,Adroja2018,Smidman2018}. The results from the heat transport~\cite{Huang2019}, optical spectroscopy
~\cite{Xu2019}, and ARPES~\cite{WuArpes} measurements support the nodeless picture. Thus, this issue requires more in-depth investigations in the future.
Perhaps previous studies on the Co-doped 122 system can be used as a reference. In the case of Ba(Fe$_{1-x}$Co$_{x}$)$_{2}$As$_{2}$, nodes
can only be detected from the heat transport measurements when the heat current is parallel
to the $c$ axis of the crystal~\cite{Reid2010}.

The fitting parameters $\alpha(H)$ and $\gamma$ are shown in Figs. 3(b) and (c). The value of $\alpha(H)$ shows a clear decrease
with an increase in the magnetic field up to 9 T, which is similar to that observed in the overdoped 122 system~\cite{Mu2011,Mu2015}.
Under zero field, a residual term $\gamma_0 \equiv \gamma (0)\approx$ 5.3 mJ/mol K$^2$ was revealed. Because there are 4 Fe in one f.u.,
this value is comparable to that of the 122 system~\cite{Mu2009,BaCo2010}.
Typically, this term is attributed to the
non-superconducting fraction of the sample~\cite{Chu2019} and/or the residual
quasiparticle DOS in the SC materials with nodes or S$^\pm$ gap
symmetry~\cite{BaCo2010,Wen2004}. Because the SC volume fraction of
approximately 100\% was confirmed by the magnetization measurement,
the former possible origin can be ruled out, and $\gamma_0$ may originate
from the residual quasiparticle DOS owing to the unconventional gap
symmetry, especially the possible existence of line nodes, in the present system.

The field-induced term $\Delta\gamma=\gamma (H)-\gamma_0$
reflects the information about the SC gap. It is difficult to describe
the field dependence of $\Delta\gamma$ using a simple formula owing to
the multi-band effect in this system~\cite{Wang2016}. Nevertheless,
qualitatively, $\Delta\gamma$ increases more quickly in the system
with a small minimum for the multi-gap case or with a highly
anisotropic gap structure, as has been observed in MgB$_2$~\cite{Bouquet2001}
and cuprates~\cite{Wen2004}. This is
because considerable quasiparticle DOS can be induced by the
magnetic field around the Fermi surface with the gap minimum. A
quick increase in $\gamma$ with the field below 3 T in our
data, as shown in Fig. 3(c), reflects this situation.
Considering the multi-band feature of this system~\cite{Wang2016},
typically there are two possibilities: (i) a large anisotropy
occurring on an individual FS sheet; (ii) a considerable difference in the
gap amplitudes between different FS sheets, at least one of which is
very small. Combined with the results that the $T^2$ term was discovered in electronic SH,
the former is the more likely scenario. Regarding the situation in the
relatively high field region above 3 T, the slow crawl of
$\Delta\gamma$ with field indicates the possible presence of larger
full gap(s) in some certain FSs, which is consistent with the multi-gap
picture.

\begin{figure}
\includegraphics[width=8cm]{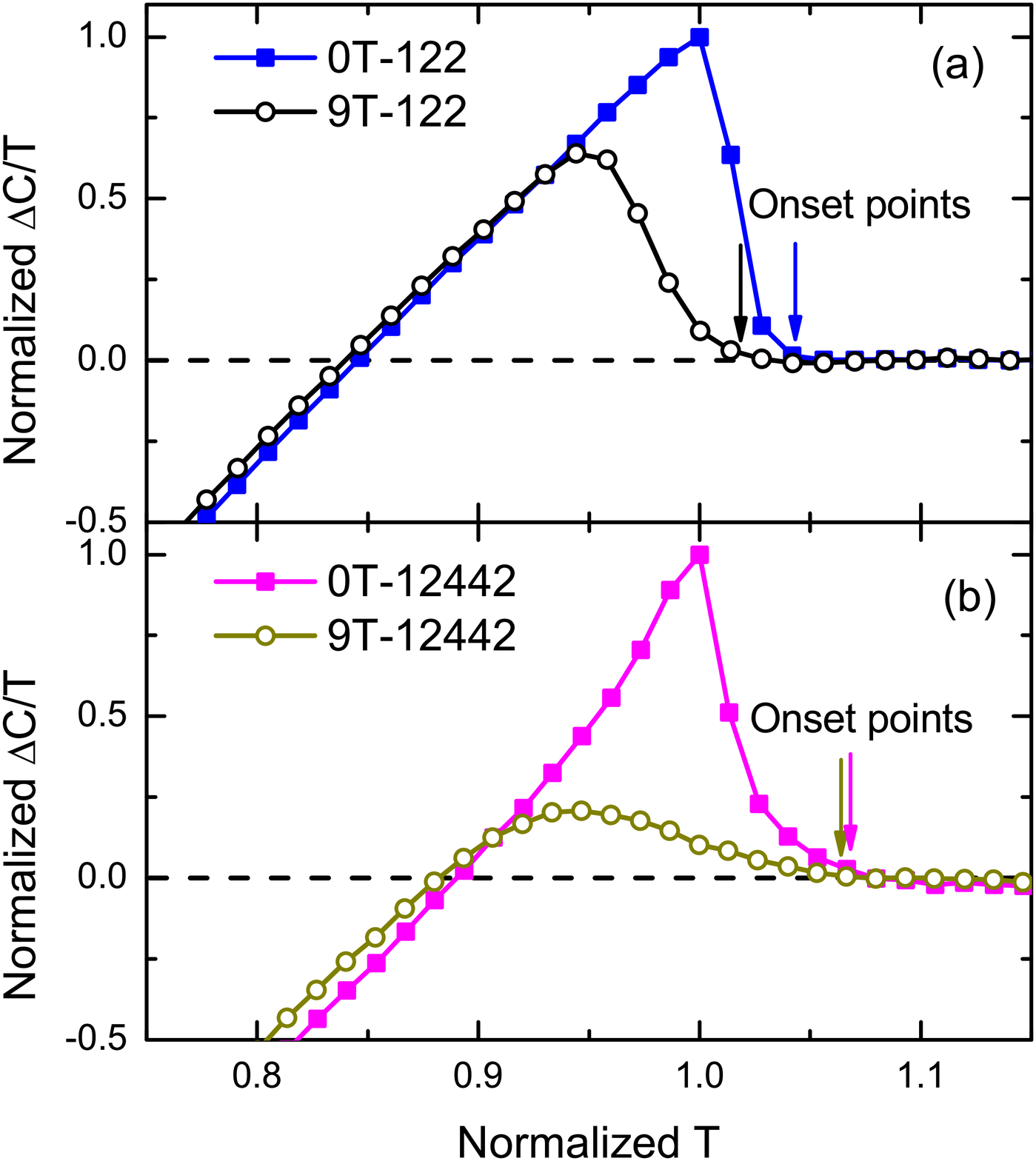}
\caption {(color online) SH data at the transition for
Ba$_{0.6}$K$_{0.4}$Fe$_2$As$_2$ (122, Ref.~\cite{Mu2009}) and
KCa$_2$Fe$_4$As$_4$F$_2$ (12442, this work). The scales of both
coordinate axes are normalized. } \label{fig4}
\end{figure}

We noticed that the quick increase in $\Delta\gamma$ is accompanied
by the considerable suppression of the SH jump at approximately $T_c$ under
magnetic field. To have a vivid impression, we plot the SH
data at approximately $T_c$ of Ba$_{0.6}$K$_{0.4}$Fe$_2$As$_2$~\cite{Mu2009}
and KCa$_2$Fe$_4$As$_4$F$_2$ in Figs. 4(a) and (b) respectively,
both of which are normalized to the position and height of the SH
peak under zero field. For the case of
Ba$_{0.6}$K$_{0.4}$Fe$_2$As$_2$, the field of 9 T only suppresses the
SH jump by 36\%. Up to 79\% of the SH jump of
KCa$_2$Fe$_4$As$_4$F$_2$ was suppressed by the same field. This
is a very large discrepancy between the two values. To have a more accurate comparison, we
need to consider the out-of-plane upper critical field $H_{c2}^c$ of
the two systems. The $T_c$ (the
slope of $H_{c2}^c$ near $T_c$, $d\mu_0H_{c2}^c/dT$$\mid$$_{T_c}$)
of Ba$_{0.6}$K$_{0.4}$Fe$_2$As$_2$ is slightly higher (lower) than
that of KCa$_2$Fe$_4$As$_4$F$_2$~\cite{Mu2009,WangCrystal-2}.
Assuming a similar evolution tendency for $H_{c2}^c$ of the two
materials toward lower temperature, the zero temperature value of
$H_{c2}^c (0)$ is slightly higher for KCa$_2$Fe$_4$As$_4$F$_2$. Thus,
the more pronounced suppression of the SH jump in this 12442 system is not
due to the difference in $H_{c2}^c (0)$ and may be the reflection of
the highly anisotropic SC gap(s).

Another notable feature is the almost unchanged onset point of the
SH anomaly under 9 T compared with the zero field data [see the
arrowed lines in Fig. 4(b)]. However, the transition temperature
in resistivity data has been clearly suppressed by the
out-of-plane field~\cite{WangCrystal-2}. It is important because
within the BCS picture, this onset point in SH should shift
monotonously to lower temperature, just as that observed in
conventional superconductors~\cite{Mu2007,Wen2009}. Compared
to the 122 system Ba$_{0.6}$K$_{0.4}$Fe$_2$As$_2$ [see Fig. 4(a)],
this feature is also noteworthy. Specifically, the field of 9 T suppresses the onset
point by more than 0.4\% for KCa$_2$Fe$_4$As$_4$F$_2$, while this
value is as high as 2.4\% for Ba$_{0.6}$K$_{0.4}$Fe$_2$As$_2$. This abnormal behavior has been reported in cuprates and has attracted
considerable attention in the
1990s~\cite{Junod1994,Alexandrov1997,Junod1999}. With the effort
from both the experimental and theoretical sides, this was
interpreted by a crossover from BCS-like superconductivity for weak
coupling to BEC-like superconductivity for strong
coupling~\cite{Junod1999}. The variable that controls such crossover was identified as $k_F\xi$~\cite{Pistolesi1994} and
sometimes as $\Delta/E_F$ for convenience~\cite{Kasahara16309},
where $k_F$ is the Fermi electron wave number, $\xi$ is the coherence
length, $\Delta$ is the SC gap, and $E_F$ is the Fermi energy. The
unexpected high values of $d\mu_0H_{c2}^c/dT$$\mid$$_{T_c}$ and
$d\mu_0H_{c2}^{ab}/dT$$\mid$$_{T_c}$~\cite{WangCrystal-2} may result
in a small coherence length $\xi$. The recent ARPES
results~\cite{WuArpes} have shown that the bottoms of the
electronic-type bands barely touch the Fermi level and form a small
electron-like elliptical Fermi pocket, which indicates the very small
amplitude of $k_F$ and $E_F$, although the detailed values were not
provided. Meanwhile, the SC gap in the electronic-type $\delta$ FS
is as large as 5.3 meV. Thus, at least within the electronic FS, we
can expect a small value of the parameter $k_F\xi$ and a relatively
large $\Delta/E_F$. We speculated that this is the internal reason
for the departure from the BCS limit and tending to the BEC limit of
this system. A more accurate estimation and analysis are needed in the
future to further confirm our speculation.

\section*{4 Conclusions}
In summary, we studied the low-temperature SH of the
12442-type KCa$_2$Fe$_4$As$_4$F$_2$ single crystal. We observed a clear
SH jump with the height of 130 mJ/mol K. The electronic SH
coefficient $\Delta\gamma$ shows a fast increase with field in
the low-field region below 3 T. Considering the presence of the $T^2$ term in the SH data, the line nodal gap structure is a very large possibility in the 12442 system.
This conclusion is further supported by the severe suppression of the SH
jump by the magnetic field. Moreover, the onset point of the SH jump
is not affected by the field, which is inconsistent with the BCS
picture and locates the present system on the crossover from the BCS
to BEC limit.

\begin{acknowledgments}
This work is supported by the Youth Innovation Promotion Association of the Chinese
Academy of Sciences (No. 2015187), the Natural Science Foundation of China
(No. 11204338), and the ``Strategic Priority Research Program (B)" of
the Chinese Academy of Sciences (No. XDB04040300). W.L. also acknowledges the start-up funding from Fudan University.
The authors would like to thank Enago (www.enago.cn) for the English language review.
\end{acknowledgments}


\end{document}